\documentclass[aps,prl,twocolumn,groupedaddress]{revtex4}
\usepackage{graphicx}
\begin{document}
\title{Unraveling quantum Hall breakdown in bilayer graphene with scanning gate microscopy}

\author{M. R. Connolly$^{1,2}$, R. K. Puddy$^1$, D. Logoteta$^3$, P. Marconcini$^3$, M. Roy$^4$, J. Griffiths$^1$, G. A. C. Jones$^1$, P. Maksym$^4$, M. Macucci$^3$, C. G. Smith$^1$}

\affiliation{$^1$Cavendish Laboratory, Department of Physics, University of Cambridge, Cambridge, CB3 0HE, UK}

\affiliation{$^2$National Physical Laboratory, Hampton Road, Teddington TW11 0LW, UK}

\affiliation{$^3$Dipartimento di Ingegneria dell'Informazione, Universit\'a di Pisa, Via G. Caruso 16, I-56122 Pisa, Italy}

\affiliation{$^4$Department of Physics and Astronomy, University of Leicester, University Road, Leicester, LE1 7RH, UK}

\date{\today}

\begin{abstract}

We use low-temperature scanning gate microscopy (SGM) to investigate the breakdown of the quantum Hall regime in an exfoliated bilayer graphene flake. SGM images captured during breakdown exhibit intricate patterns of ``hotspots'' where the conductance is strongly affected by the presence of the tip. Our results are well described by a model based on quantum percolation which relates the points of high responsivity to tip-induced scattering between localized Landau levels.

\end{abstract}

\maketitle

Quantized plateaus in the Hall conductance of a two-dimensional electron system (2DES) develop whenever the Fermi level is in a gap between two Landau levels (LL), making them ideal spectroscopic markers for exploring the sensitivity of LL energy spectra to a wide range of degeneracy-breaking interactions. The plateau structure of single- and few-layer graphene, for instance, has revealed that valley, spin, and sublattice degeneracies are broken by changing the morphology and topography of the crystal lattice, the length scale and strength of the potential disorder landscape \cite{Zhang2006,Novoselov2006,Zhao2010}, and the type of layer stacking order \cite{Taychatanapat2011}. Microscopically, quantum Hall effect (QHE) plateaus occur because electrons in the bulk follow closed and thus localized paths, while current-carrying extended states run along the free edges where they are protected from back-scattering and dissipation \cite{Chalker1988}. The QHE breaks down in the transition regions between plateaus because electrons percolate through a network of bulk states, leading to back-scattering between edge states and a non-integral contribution to the Hall voltage. Since the resolution, length, and quantization accuracy of the plateaus are governed by the topological properties of this network, understanding the microscopic details of QHE breakdown in graphene devices will be key to exploring interactions in finer detail, and for implementing graphene as a metrological standard of resistance that can be operated at lower magnetic fields and higher temperatures \cite{Tzalenchuk2010, Janssen2011, Giesbers2008}.

\begin{figure}[!t]
\includegraphics{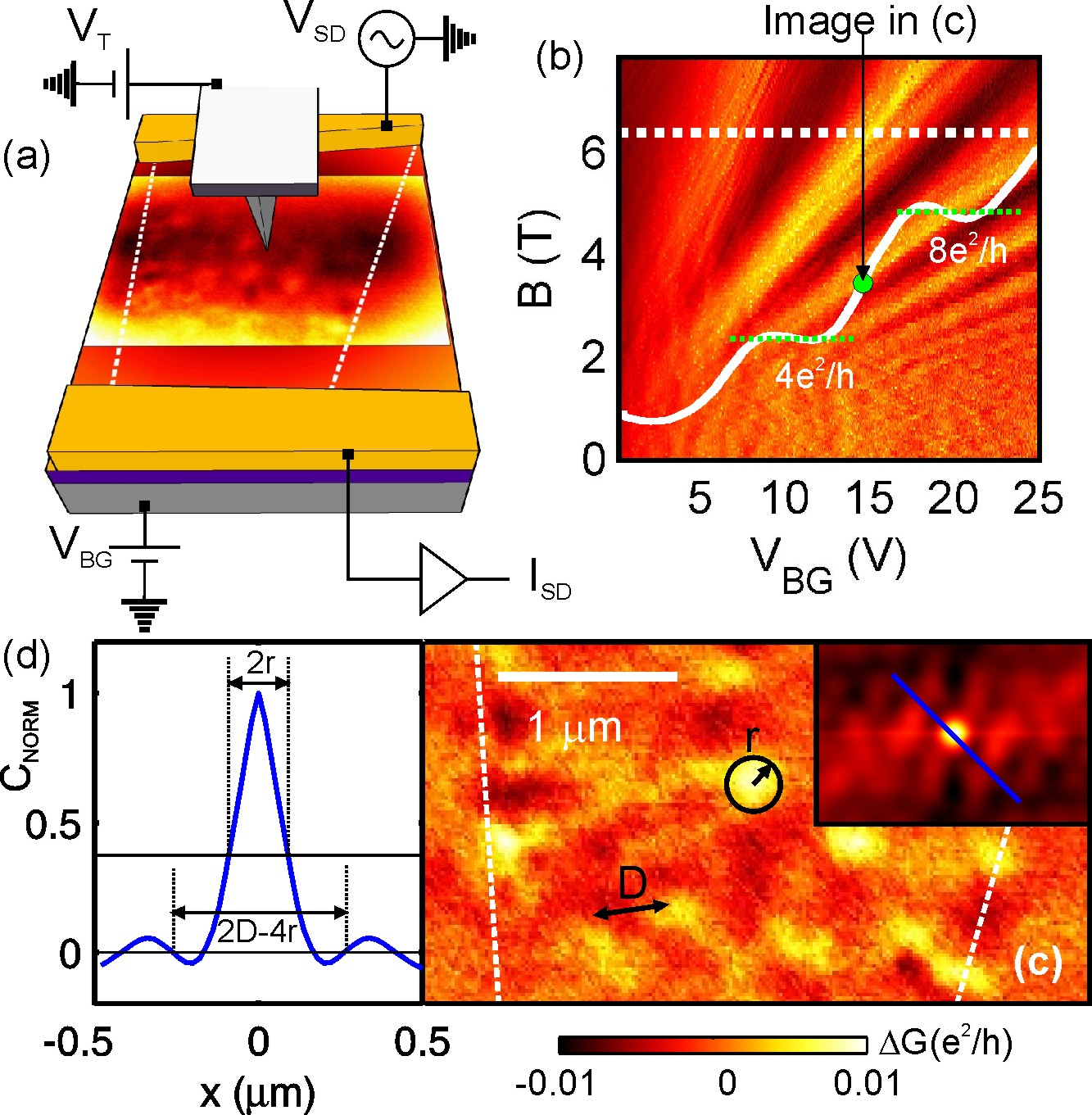}
\caption{(a) Circuit used to perform scanning gate microscopy. The edges of the flake are indicated by white dashed lines and superimposed over the surface is a raw SGM image taken at a lift height of 50 nm. (b) Landau level fan diagram showing the numerical derivative of the conductance as a function of back-gate voltage and magnetic field at $T$$\approx$8 K. The source-drain bias voltage is V$_{SD}$=1 mV. The white (solid) line is a trace of the conductance along the dashed line at $B$=6.2 T, showing plateaus at filling factors of 4 and 8. (c) Flattened SGM image obtained with the tip at $\approx$50 nm lift-height. The white dashed lines indicate the edge of the flake. Inset: Autocorrelation function taken over the flake. (d) Line profile through the peak of the 2D autocorrelation function. Labeled length scales $r$ and $D$ correspond to the radius and separation between the conductance hotspots, respectively.}    
\label{Fig:Fig1}
\end{figure}
 
In this letter we use scanning gate microscopy (SGM) to unravel the paths of electrons during QHE breakdown and show that transport is well described by quantum percolation between localised LLs. Although the nature of quantum Hall localization in graphene has enjoyed much attention recently \cite{Miller2010, Jung2011, Martin2009, Luican2011, Rutter2011}, the topological origin of the QHE breakdown has previously only been examined using SGM in GaAs sub-surface 2DESs \cite{Woodside2001, Baumgartner2007, Kicin2004}. Both SGM results and those obtained by less invasive techniques such as scanning force microscopy \cite{Ahlswede2001} and scanning tunnelling microscopy \cite{Hashimoto2008} were well described within a single-particle framework \cite{Dubi2006}.

We investigate a graphene flake (dimensions $\approx$ 2.5 $\times$ 6 $\mu$m$^{2}$) mechanically exfoliated from natural graphite onto a highly doped Si substrate capped with a 300 nm thick SiO$_{2}$ layer. The flake was identified as a bilayer from its optical contrast \cite{Blake2007}, and two (5 nm/30 nm) thick Ti/Au contacts were patterned using e-beam lithography, thermal evaporation, and standard PMMA lift-off processing. Figure \ref{Fig:Fig1}(b) shows the numerical derivative of the two-terminal conductance of the device as a function of back-gate voltage $V_{BG}$ and magnetic field $B$ at a temperature $T$$\approx$8 K ($V_{SD}$= 1 mV). As anticipated for two-terminal bilayer graphene devices, N-shaped conductance plateaus quantized in units of 4$e^2/h$ develop as a result of edge channel conduction and strong localization in the QH regime \cite{Falko2007, Williams2009}. 

To probe the QH state locally during breakdown, we tune the conductance to a value between the first and second quantized plateaus ($V_{BG} = 17$ V, $B=$ 6.2 T), and image the device using SGM (see Refs. \cite{Crook2000,Topinka2000,Baumgartner2007, Connolly2010, Berezovsky2010a,Berezovsky2010b, Jalilian2011} for more details). In brief, SGM involves scanning a sharp metallic tip over the surface of graphene while measuring its conductance. A schematic of our SGM setup is shown in Fig. \ref{Fig:Fig1}(a). Our scanning probe microscope head (AttoAFM I) is mounted to the mixing chamber of a dilution refrigerator. The oscillation of the cantilever is measured using standard interferometric detection with a fibre-based infra-red laser. We use a Pt/Ir coated cantilever (NanoWorld ARROW-NCPt) with a nominal tip radius of 15 nm. In order to avoid any cross-contamination between the tip and the flake during SGM, once the flake is found using tapping mode, we switch to lift-mode with the static tip at a lift-height of $\approx$50 nm. As a precaution against drift when scanning close to the 100 nm-thick metallic contacts, we measure the conductance using an RF lock-in amplifier with an excitation frequency matched to the resonant frequency of the cantilever over the bare SiO$_2$. The stray field from the metallic contacts is sufficient to excite the cantilever into oscillation over the SiO$_2$, though over the graphene the cantilever is off-resonance and static. In order to obtain a good signal-to-noise ratio at these lift heights, we use an excitation voltage of 10 mV. While this is rather large for low temperature transport measurements, it is still less than the energy separation ($\approx$50 meV) between consecutive LLs at the magnetic fields and filling factors examined in our experiments.  

A typical scanning gate micrograph is shown in Fig. \ref{Fig:Fig1}(c). A striking feature of the image is a texture consisting of $\approx$ 100 nm-sized ``hotspots'' where the conductance is strongly modulated by the tip. Note that this fine structure appears against a broad background modulation, which probably stems from the long-range gating effect of the tip cone \cite{Ouisse2008}. To examine just the fine pattern in more detail, the image in Fig. \ref{Fig:Fig1}(c) was flattened by subtracting a parabolic background with image analysis software \cite{Horcas2007}. To analyze features in the resulting image quantitatively, we calculate the two-dimensional autocorrelation function $C(x,y)$ shown in the inset of Fig. \ref{Fig:Fig1}(c). Owing to the roughly uniform density and regular size of the hotspots, $C(x,y)$ exhibits oscillations with periodicity governed by the average hotspot spacing ($D$) and a peak close to zero whose half-width reflects their size ($r$) \cite{Heilbronner1992}. Figure \ref{Fig:Fig1}(d) shows a section of $C(x,y)$ taken along the blue line in the inset of Fig. \ref{Fig:Fig1}(c), allowing us to make estimates for $r$$\approx$ 90 nm and $D$$\approx$ 450 nm.  

\begin{figure}[!h]
\includegraphics{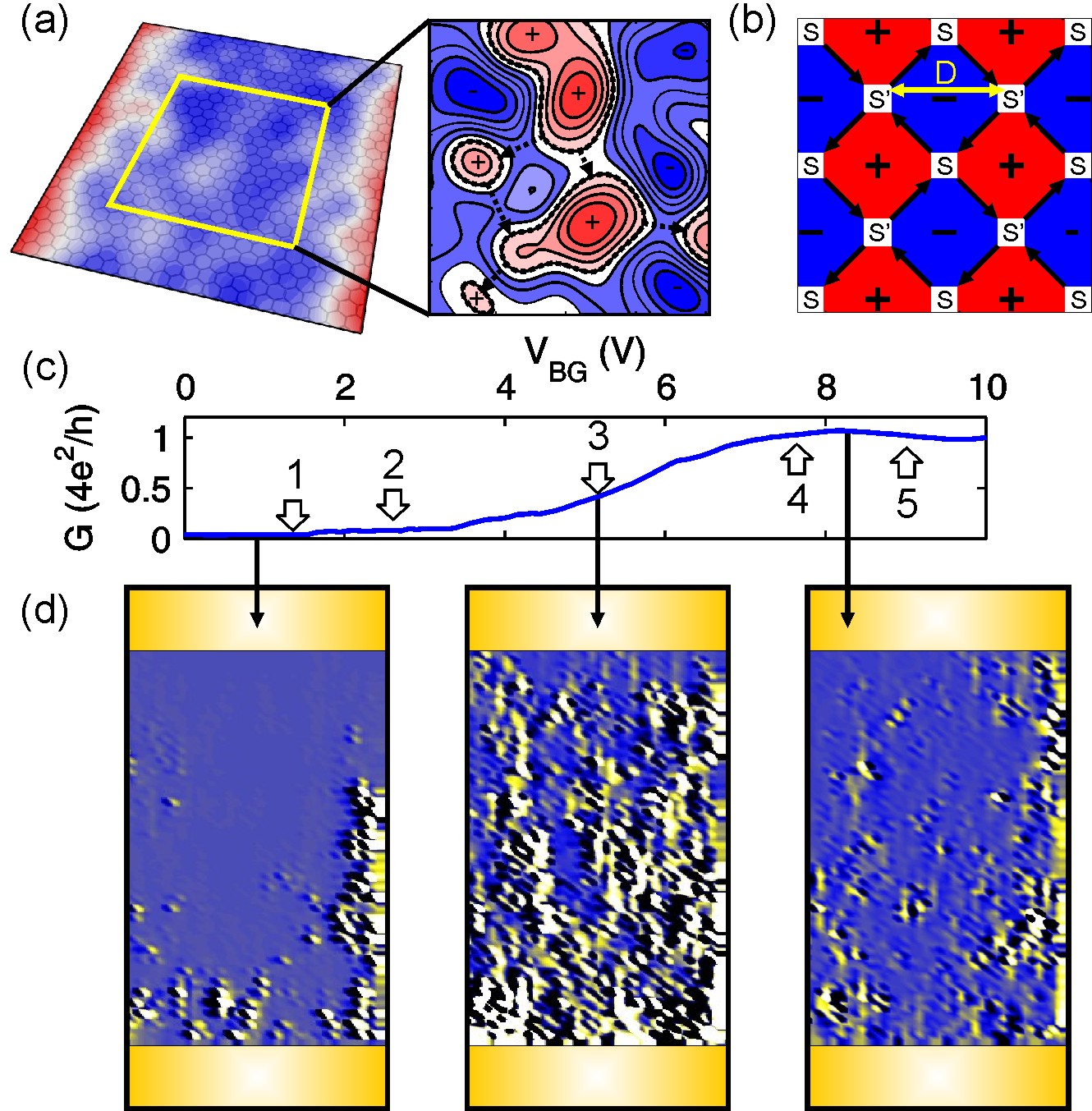}
\caption{(a) Depiction of a disorder potential in graphene. At the Landau level centre, electrons circulate around hills (+) and valleys (-) in the disorder potential, coming into tunnelling proximity at the saddle points. An illustrative conducting orbital is depicted as a dashed contour. (b) Projection of the system into a network model characterized by the scattering matrices $S$ and $S'$ defining the tunnelling probability across the nodes. (c) Conductance of a network consisting of 38 $\times$ 76 nodes as a function of backgate voltage. The total conductance is obtained by averaging over a range of Fermi levels corresponding to a bias window of $\approx$ 10 meV. Numbers refer to the corresponding simulated images in Fig. \ref{Fig:Fig3}. (d) Maps of the current density numerically obtained for the three indicated values of back-gate voltage.}    
\label{Fig:Fig2}
\end{figure}

To understand the origin of these hotspots and how they relate to the underlying electron trajectories, we adopt a phenomenological network model proposed by Chalker and Coddington \cite{Chalker1988, Kramer2005}. This method has been successfully used to study percolative transport in QHE and, incidentally, has also been shown to map, with the proper positions, onto the two-dimensional Dirac equation \cite{Ho1996, Snyman2008}, governing electron dynamics in monolayer graphene \cite{Marconcini2011}. This model assumes that conduction through the flake around the LL can only occur along a path of connected localized states. Here we employ the conventional picture that electrons perform cyclotron orbits while drifting along equipotential contours in the electrostatic landscape \cite{Chalker1988}, and it is these orbits which we refer to as localized states. The tip can thus only affect the conductance by perturbing the potential of the saddle points where two such states approach each other and tunneling becomes possible [Fig. \ref{Fig:Fig2}(a)]. For simplicity, our network consists of a regular array of saddle points, as shown in Fig. \ref{Fig:Fig2}(b) \cite{Chalker1988}. As in Ref. \cite{Kramer2005}, the transmission of each node of the network is parametrized by means of a dimensionless quantity $\theta$, which is a nondecreasing function of the difference between the energy of the incident electron and the potential of each saddle point. We decompose the potential associated with each saddle point into two components: the value of the potential in the absence of back-gate and probe bias, and the perturbation due to biasing these electrodes. This last component thus links the value of $\theta$ to the voltage applied to the back-gate and the tip. Disorder is introduced by randomizing the values of the potential in the absence of bias, and keeping the associated variance as a free parameter provides a way to extract the size of the potential fluctuations in a real sample. Further disorder is included by randomizing, as in the model of Chalker and Coddington, the phase shifts associated with each link, to take into account the random relative positions of the saddle points. 

Figure \ref{Fig:Fig2}(c) shows the transmission of a sample network as a function of $V_{BG}$ with the tip voltage held at zero, showing a gradual transition from an insulating to a fully transmitting condition, in good agreement with the conductance versus $V_{BG}$ measured experimentally in Fig. \ref{Fig:Fig1}(b). At low values of $V_{BG}$, corresponding, on average, to low values of $\theta$, the probability of tunneling between localized states is small due to the large separation between neighbouring localized states residing in the bottom of potential valleys, and the transmission is thus suppressed. At high values of $V_{BG}$, corresponding, on average, to large values of $\theta$, transmission is perfect owing to the formation of completely conducting links along the edge of the network. At intermediate values of $V_{BG}$, transmission increases due to conduction through a series of localized states tunnel coupled at saddle points. Representative maps of the current density during these stages are shown in Fig. \ref{Fig:Fig2}(d). The network model thus provides an ideal framework for simulating SGM images: we perturb the $\theta$ value at each node and plot the change in conductance $\Delta G$ as a function of node position. The amplitude of the perturbation at each saddle point is obtained assuming that the probe produces a Lorentzian contribution to the potential centered on the tip. We performed numerical simulations to determine the effect of a realistic tip geometry on a graphene bilayer sheet and found the perturbation is well described by the sum of two Lorentzians. We use the narrower of the two, which has a half width at half maximum of 50 nm \cite{Connolly2011b}, to describe the local extent of the tip. A hotspot implicates a particular node group as part of a current path that makes a substantial contribution to the conduction of the whole network, mimicking the experimental situation precisely. We note here that  several studies \cite{Steele2005, Ilani2004, Yacoby1999,Martin2009} have suggested the need to consider many-body effects within the bulk of samples possessing small-lengthscale/large-amplitude potential fluctuations. While we do not include this effect in our simulations, such many-body effects are not incompatible with the Chalker-Coddington model and may in fact be introduced via a Fermi-level dependence of the saddle-point potential \cite{Ho1999}.

\begin{figure*}
\includegraphics{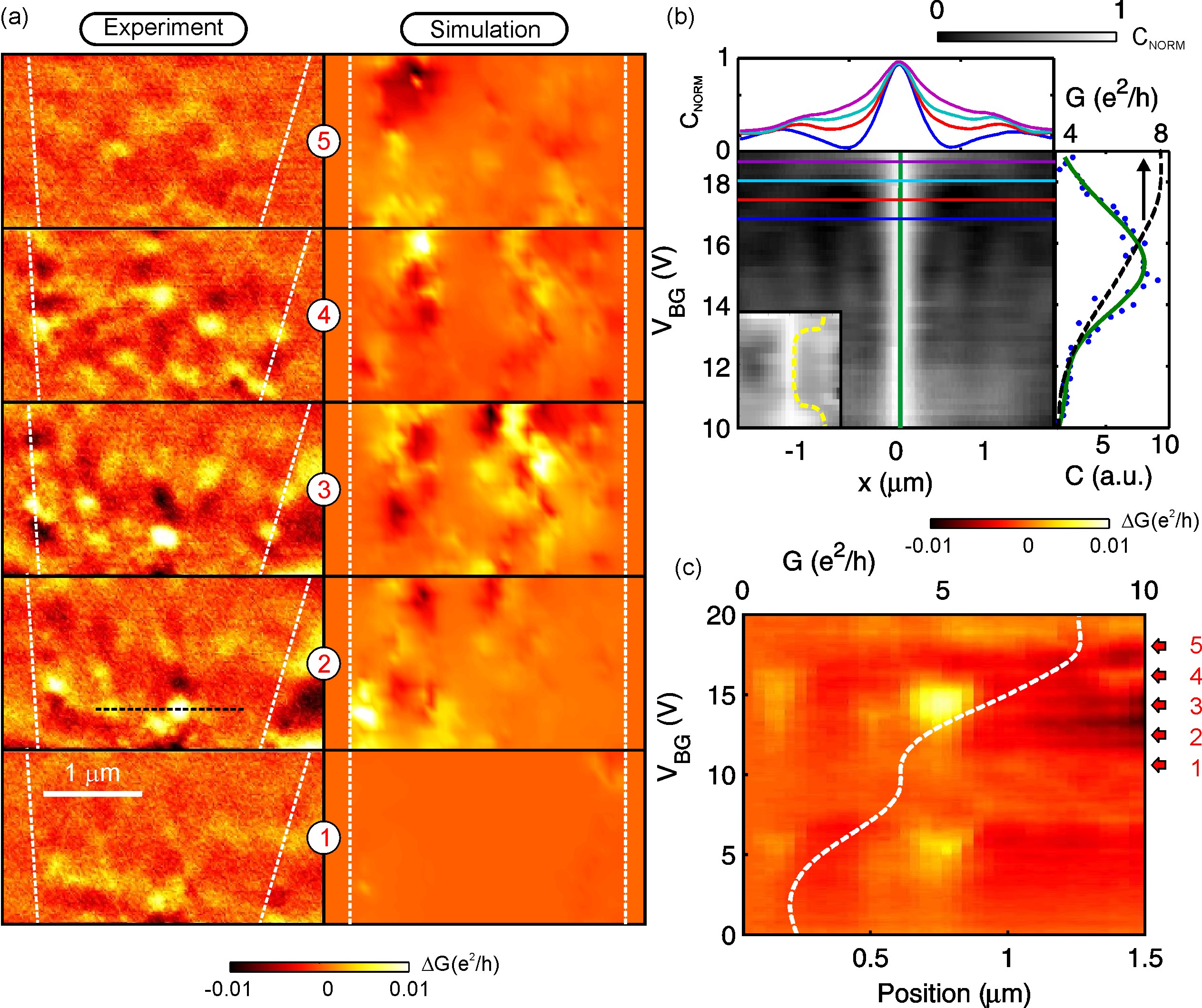}
\caption{(a) Sequence of SGM images captured at different back-gate voltages along the riser between the first and second quantized conductance plateaus. Adjacent are simulated images of the change in conductance obtained at the back-gate voltages indicated by the numbered arrows in Fig. \ref{Fig:Fig2}(c). White dashed lines indicate the edge of the flake. Black dashed line in image 2 shows the line along which the profile shown in (c) was taken. (b) Cross section of the 2-D normalized autocorrelation function as a function of back-gate voltage. Plots above the main figure show cross sections at back-gate voltages indicated by lines of the same color. To enable a comparison between the autocorrelation function and the corresponding bulk transport, the conductance and the amplitude $C$ of the central peak are plotted as a function of back-gate voltage to the right of the main figure. Data are shown by blue points and the green curve is a guide to the eye. Inset: autocorrelation function obtained from the simulated images in Fig. \ref{Fig:Fig4}. The yellow dashed line indicates the edge of the central peak. (c) $\Delta$$G$ as a function of position and back-gate voltage along the line in image 2 spanning two hotspots. Overlaid is the conductance as a function of back-gate voltage. Numbered arrows indicate the back-gate voltages at which the corresponding images in (a) were captured. }    
\label{Fig:Fig3}
\end{figure*}

To test this picture for quantum Hall breakdown, we monitor the evolution of the hotspots with back-gate voltage at a fixed magnetic field of 6.2 T. Figure \ref{Fig:Fig3}(a) shows a sequence of SGM images captured at different values of $V_{BG}$ as the conductance increases on the riser between the first and second quantized plateaus. A casual inspection of the images shows that on the plateaus themselves (images 1 and 5) the image texture is characterized by weak long-range fluctuations in $\Delta G$, while on the riser it becomes more intricate and the intensity of individual hotspots increases. The hotspot intensity also appears most pronounced when the LL is half filled (image 3). This behaviour is concisely represented and more detail is revealed in Fig. \ref{Fig:Fig3}(b), which shows the evolution of the normalized autocorrelation function with back-gate voltage. The data are extracted from a set of 50 images taken between back-gate voltages of 10 and 20 V in 0.2 V increments. The width of the peak around zero corresponding to 100 nm-sized hotspots [c.f. Fig. \ref{Fig:Fig1}(c)] remains constant on the riser, diverging to around 1 $\mu$m at either end where the flake enters the QH regime. This trend is clearly depicted in Fig. \ref{Fig:Fig3}(b), where we show several line-profiles at different back-gate voltages at the edge of the plateau. To determine whether these features of our data are peculiar to the transition between the $\nu=4$ and $\nu=8$ states (where $\nu$ indicates the filling factor), we captured a similar set of images between the Dirac point and the $\nu$=4 plateau. The result is summarized in Fig. \ref{Fig:Fig3}(c), which shows a $\Delta G(x)$ profile across several hotspots [see image 2, Fig. \ref{Fig:Fig3}(a)] as a function of back-gate voltage spanning both the first two risers. The hotspots appear and reach their peak intensity at the same position along both risers, confirming that the observed image sequence is robust and is controlled by the filling factor of the top LL relative to half-filling. Note that the average intensity of all the hotspots is also reflected by the amplitude of the central autocorrelation peak, which also reaches a maximum at the middle of the riser [see Fig. \ref{Fig:Fig3}(b)].   

The right panel of Fig. \ref{Fig:Fig3}(a) shows a representative sequence of simulated SGM images taken at the values of $V_{BG}$ marked in Fig. \ref{Fig:Fig2}(c). Key experimental observations such as the increasing intensity and intricacy of the texture towards the centre of the LL are well captured by the simulations. In particular the inset of Fig. \ref{Fig:Fig3}(b) shows the simulated autocorrelation function, which displays the same divergence of hotspot size towards the ends of the risers, and the amplitude of the conductance variations are also in good quantitative agreement. Thus our experimental observations are naturally explained within the proposed framework as arising from the increased likelihood of tip-enhanced tunneling due to the increased proximity of bulk states at the centre of the LL, as illustrated in Fig. \ref{Fig:Fig2}(d). In our calculations we have considered different values for the mesh size, i.e. the separation $D$ between saddle points (or nodes), from 50 to 100 nm, and we settled on the value of 60 nm, which appeared to yield the best agreement with the experimental results. The considered potential fluctuations are uniformly distributed around zero, as in Ref. \cite{Dubi2006}, within an interval of amplitude $\simeq$ 10 meV. Both the spatial extent (60 nm) and strength of the disorder which adequately reproduce the experimentally determined values for $r$ and $D$ are slightly larger than, though in reasonable agreement with, values existing in the literature \cite{Rutter2011,Deshpande2009}.

\begin{figure}[!t]
\includegraphics{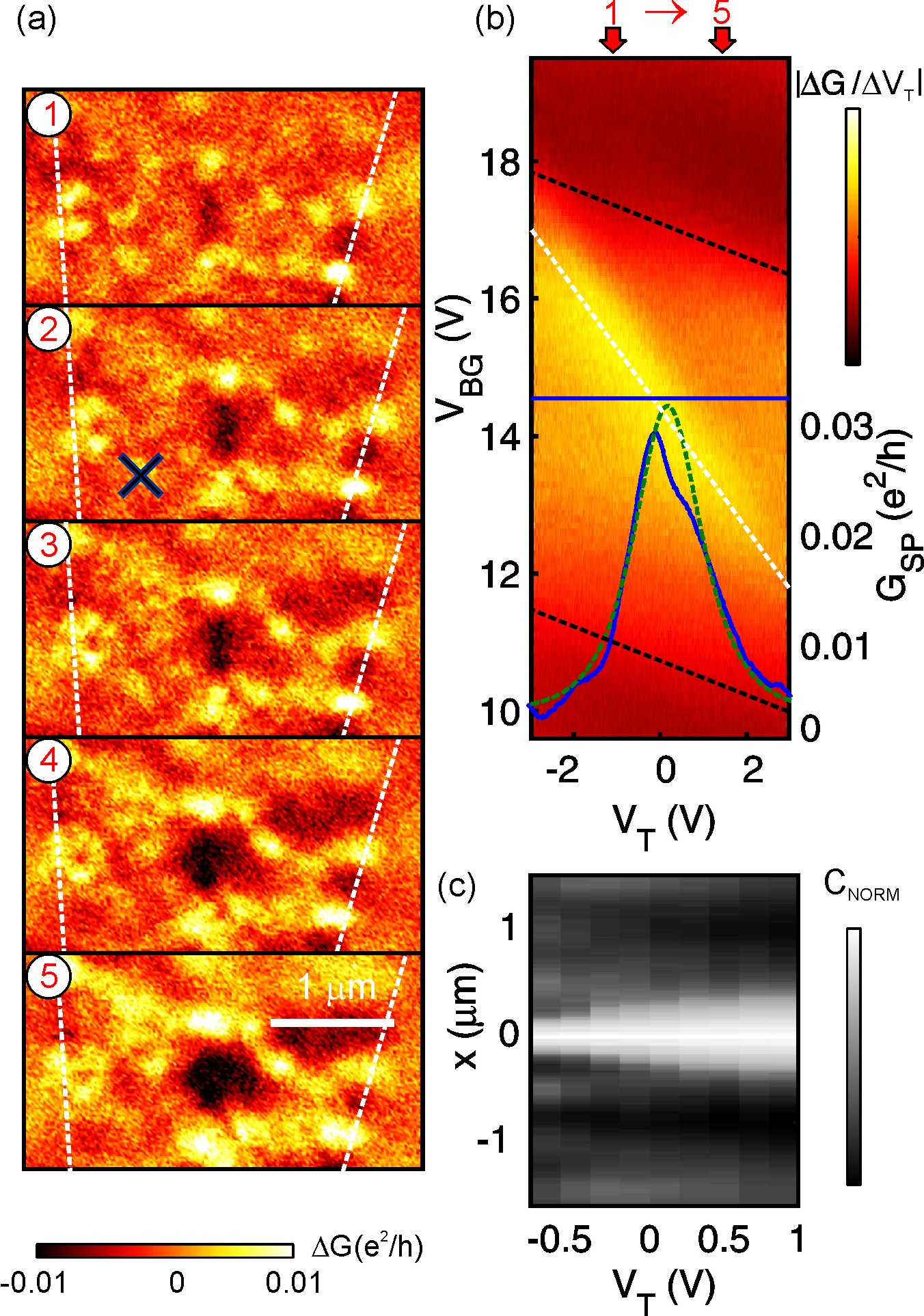}
\caption{(a) Sequence of SGM images captured at a fixed back-gate voltage $V_{BG}$= 7.4 V with tip voltages of -0.8 V (image 1) to 0.8 V (image 5) in 0.4 V steps. (b) Numerical derivative of the conductance with respect to the tip voltage as a function of back-gate and tip voltage. Black and white dashed outlines indicate the edges of the riser between quantum Hall plateaus and the conductance peak due to the hotspot, respectively. The experimental (solid) and theoretical (dashed) line profiles superimposed show the conductance as a function of tip voltage with the tip parked over the hotspot indicated by a cross in image 2 of (a). A linear slope has been subtracted to remove the background gating effect of the tip-cone. (c) Normalized autocorrelation profile as a function of tip voltage.} 
\label{Fig:Fig4}
\end{figure}

To obtain further insight into the properties of individual hotspots, Fig. \ref{Fig:Fig4}(a) illustrates a typical sequence of images captured with increasing tip voltage ($V_T$) at a fixed $V_{BG}$=7.4 V, midway along the riser between the Dirac point and the $\nu$=4 plateau where the hotspots are well defined. The size of each hotspot tends to increase with increasing tip voltage and they appear to merge into connected areas. This behaviour is clearly reflected in the evolution of the autocorrelation function shown in Fig. \ref{Fig:Fig4}(c). $C(x)$ exhibits oscillations at low $V_T$ due to the presence of isolated hotspots. At higher $V_T$ the central peak broadens as the hotspots increase in size and the oscillations decay and merge together. 

We find that the intensity of each hotspot displays a more subtle behaviour which also depends on $V_{BG}$. This is illustrated in Fig. \ref{Fig:Fig4}(b), which was acquired by parking the tip over the hotspot marked in image 2 of Fig. \ref{Fig:Fig4}(a), and sweeping $V_T$ at different values of $V_{BG}$. As expected from the long range gating effect, changing the tip voltage shifts the overall position of the riser in back-gate voltage (black dashed lines). The presence of the hotspot under the tip causes a narrower peak in the conductance to move through the riser with a steeper slope in the $(V_T,V_{BG})$ plane (white dashed line). As mentioned previously, the perturbation from the entire cantilever can be well described by the sum of two Lorentzians, one broad and shallow and the other narrow and deep. Hence the difference in these slopes can be understood in terms of the difference in capacitive coupling from the tip-apex relative to the tip-cone.

The typical form of this peak in saddle-point conductance, $G_{SP}(V_T)$, is shown superimposed in Fig. \ref{Fig:Fig4}(b), where we have subtracted a linear function with slope $\beta$ from the raw data in order to eliminate the gating effect of the tip-cone ($G_{SP}(V_T) = G(V_T)-\beta V_T$). The full-width of the peak in $V_T$ is $\approx$ 2 V and at its maximum the conductance increases by a few percent of $e^2/h$. To understand the particular form of this data we draw from a theoretical model for the transmission of 1-D channels across a saddle point. In the quantum Hall regime the conductance of a saddle point is given by $$G_{SP} = \frac{e^2}{h}\frac{P}{1+P^2},$$ where $e$ is the modulus of the elementary charge, $h$ is Planck's constant, and $P$ is the ``backscattering'' parameter \cite{Oswald1998}. The symmetry of $G_{SP}$ about $V_{SP}$ demands that $P$ is an exponential function of the relative filling factor. For an ``eggbox'' potential it takes the simple form $P=\exp(\pm\frac{D^2E_F}{e\tilde{V}}\frac{eB}{h})$, where $D$ is the separation between saddle points, and $\tilde{V}$ is the mean fluctuation in the disorder potential. The green dashed curve in Fig. \ref{Fig:Fig4} shows a good fit of the data to this expression, assuming the node separation $D\approx$60 nm deduced from our numerical simulations.

In conclusion, we have examined the breakdown of the quantum Hall effect using low temperature scanning gate microscopy and numerical simulations. In the quantum Hall regime, the position of the scanning probe tip has a weak influence on transport because conduction occurs at the edges while bulk localized states are well isolated from each other. During quantum Hall breakdown we found that the conductance is strongly modulated by the tip at specific locations, and these conductance ``hotspots'' were found to repeat at the same relative filling factor. To understand our experimental observations we performed numerical simulations based on a network model for percolation between localized states. By comparing the divergence of the autocorrelation function at the edges of the riser with the simulation we were able to optimise the network parameters, yielding a 60 nm node separation and disorder fluctuation of $\approx$ 10 meV, both in good agreement with previous studies. Finally, by imaging at different tip potentials we find that the conductance modulation at individual hotspots is well described a by theoretical model assuming that transmission occurs via percolation of 1-D channels across individual saddle points. Our results demonstrate that SGM is a powerful tool for probing the quantum Hall state in graphene and provides an important insight into the interaction between potential disorder and magnetic field induced localization.

This work was financially supported by the European GRAND project (ICT/FET, Contract No. 215752). 


\end{document}